\documentstyle[11pt,newpasp,twoside,epsf]{article}
\def\edcomment#1{\iffalse\marginpar{\raggedright\sl#1\/}\else\relax\fi}
\reversemarginpar

\begin{document}
\title{Searching for Jets in Asymmetrical Nebulae with the Hubble Space Telescope}
 \author{J. Patrick Harrington}
\affil{Department of Astronomy, University of Maryland, College Park, Maryland 20742}
\author{Kazimierz J. Borkowski}
\affil{Department of Physics, North Carolina State University, Raleigh, North Carolina 27695}

\begin{abstract}
The jets seen in NGC 6543 motivated us to search for similar features
in other PNe. Our HST snapshot program looked for jets using the [N II] 
and H$\alpha$ narrow-band filters. Although spectacular jets are found in
proto-PNe, true jets seem rare among mature PNe. In the later group, 
IC 4593 is the best case in our sample. We distinguish between jets and
a number of interesting ``jet-like'' features, e.g., ``cometary'' structures
with a dense globule at the end facing the central star. 
\end{abstract}

\section{Introduction}

Observations of Herbig-Haro objects like HH30 and HH34 have provided perhaps
the best images of narrow, continuous jets. These jets are known to
originate in the accretion disks surrounding the associated T-Tauri stars.
Recently, morphological studies of PNe have revealed many ``point-symmetric''
structures, which would have a natural explanation if the nebula were once
subjected to the effects of precessing jets. But the
existence of jets seems at variance with the conventional picture of the PN
central star, which does not involve an accretion disk. In this context, any
observations of jets in PNe, which provide a more direct indication of 
stellar accretion disks, are especially interesting.

HST observations of NGC~6543 (the ``Cat's Eye Nebula'') confirmed the earlier 
indications (Miranda \& Solf 1992) of a remarkable pair of jets in this
object (Figure 1). We wanted to see if similar structures exist in other 
PNe. With this in mind, we carried out a ``snapshot'' imaging program of 
likely nebulae with the HST.

\section{The HST Snapshot Program}

\begin{figure}  
\plotfiddle{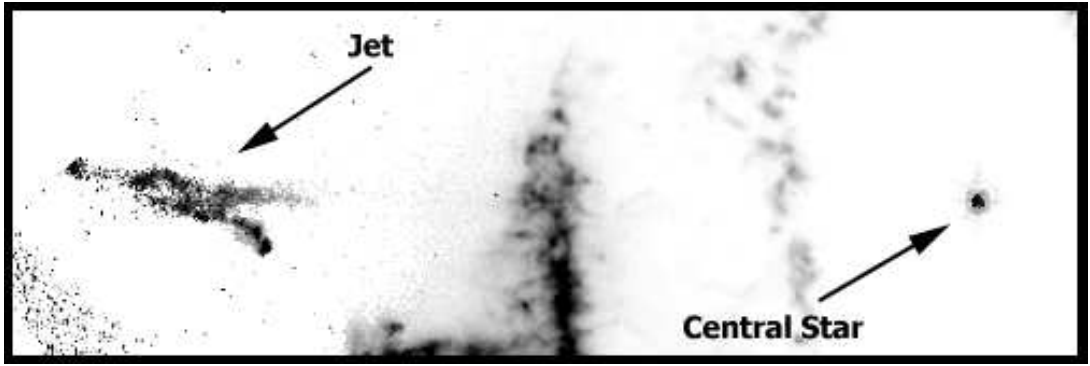}{3.8cm}{0.0}{80}{80}{-200}{0}    
\caption{One of the pair of jets in NCG 6543.}
\end{figure}

\begin{table}
\caption{Targets Observed in Snapshot Program No. 6347}
\begin{tabular}{lllll}
\tableline
 & HST Name & Other Name(s) & Filter(s) & Comments \\
\tableline
1& PK190-17D1 &J 320 & H$\alpha$ & multi-polar \\
2& PK242-11D1 &M 3-1  & [N II], H$\alpha$ & large point-symmetric PN\\
3& NGC2392  &  PK197+17D1, Eskimo &[N II], H$\alpha$ & cometary tails \\
4& IC4593  &  PK025+40D1 & [N II], H$\alpha$ & bullets and jets \\
5& NGC6210 &  PK043+37D1 & [N II]  & one bullet \\
6& PK336-05D1 & He 2-186 & [N II]  &  isolated bi-knots \\
7& PK003+02D1 & Hb 4 &[N II], H$\alpha$ & corkscrew jets, not aligned \\ 
8& He 3-1475 & IRAS 17423-1755  & [N II] & high velocity, conical jets \\
9& PK032+07D2 & PC 19  & [N II] & spiral! (point symmetry) \\
10& PK051+09D1 & Hu 2-1 &  [N II], H$\alpha$ &  twisting axis \\
11& PK032-02D1 & M 1-66 & H$\alpha$  & conical structures \\
12& PK055+02D2 & He 1-1 & H$\alpha$  & radial streaks \\ 
13& M1-92 & (Minkowski's footprint)& [N II] & PPN with bipolar outflow \\
14& NGC6881 & PK074+02D1 & [N II], H$\alpha$ & huge bipolar \\
\tableline
\tableline
\end{tabular}
\end{table}

\begin{figure}  
\plotfiddle{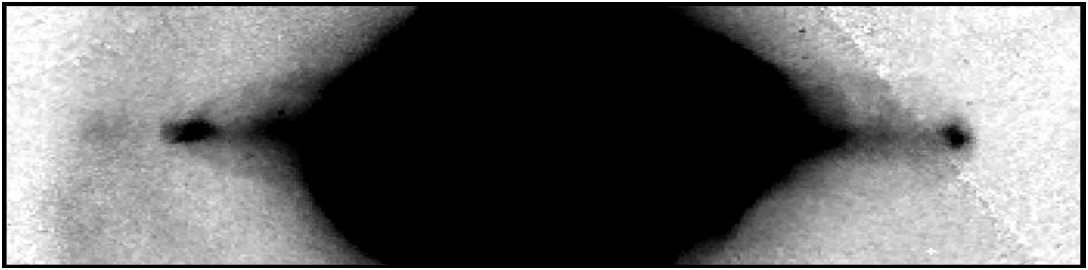}{3.5cm}{0.0}{80}{80}{-200}{0}    
\caption{The ``bullets'' and associated bow shocks in IC 4593.}
\end{figure}
 
We selected our targets based on  three criteria: (1) reports of
high velocity flows (e.g., He 3-1475, the Eskimo), (2) point-symmetric morphologies,
(3) ground-based images that showed ``jet-like'' structures. Since we found
that in the case of NGC~6543 the jets were best seen in the ratio of the
[N II] $\lambda$6584 to H$\alpha$ filter images (Figure 1 is such a ratio
image), we requested [N II] and H$\alpha$ images of our targets. Snapshot
programs limit the exposure time; all our images are 10 minute exposures.
When the program ended on 29 June 1999, 14 objects had been observed, most
in [N II], but only 6 in both filters. The nebulae we observed are listed in Table
1. Since the program was basically a morphological survey, we set the public
release time for our data at 2 months -- all are now available from the STScI 
archives.

\section{Results}

While we have found a variety of interesting radial structures in these nebulae,
we have come to the conclusion that, at least for mature PNe (as opposed to 
proto-PNe), real jets are rare. We consider a true jet to be a narrow, continuous,
high-velocity flow. There are two types of long, radial structures which we
feel have no relation to jets: we will call them {\it cometary structures} and
{\it rays}. Cometary structures have a bright ``head'' at the end nearest the star,
from which a low-ionization tail, often sinuous, extends outward. These features 
were first seen in the Helix over three decades ago.  HST images have shown
that the heads are neutral globules which are photo-evaporating. The globules
are drifting outward more slowly than the surrounding ionized gas; the tails
presumably result as evaporated material is dragged back by the flow. Of the
nebulae imaged for this program, we found that the Eskimo has an extensive set of
cometary structures. Their sinuous tails could indicate a subsonic flow past the
globules. Such comets are seen in other PNe, such as NGC 6543 and A 30 (Borkowski
et al. 1995). The tails may be quite long: the ``jet'' in NGC~7354 looks like a
comet to us. There is, however, one linear feature in the Eskimo at P.A. 90\deg~ 
which is different: very thin, straight and directed exactly away from the star.
This feature also has a clump at its head. It seems too narrow to be a jet. In  
NGC~6543 there is a bundle of such {\it rays} outside the northern cap. It seems
probable that the rays are ``shadow columns'', where the gas, shielded from the 
direct stellar radiation, is ionized and heated only by the diffuse field. Such
gas would be less highly ionized, cooler and hence denser. Such shadow columns
might only form if the gas is sufficiently quiescent.

Our images of nebulae chosen for point-symmetry (Hu 2-1, J320, M 3-1, PC 19) 
show no jet-like structures, but the symmetry is seen to be much more detailed
and precise than was apparent from ground-based observations. \mbox{Hb-4} is a curious
case. It has two jet-like structures well away from the main nebulosity, but they
are not even approximately co-axial. L\'{o}pez et al. (1997) found these structures
have velocities of $\pm$150 km/s. We would class these features as ``FLIERs''. Our 
HST images reveal that these structures have a distinct ``corkscrew'' morphology;
hints of similar structure are seen in the ansae of NGC~6543. 

Perhaps the best case for a real jet in our sample -- aside from the proto-PN
discussed below -- is IC 4593. Two ``bullets'' emerge from the main body of the
nebula on an axis directly through the central star. Trails of material are seen
connecting the bullets with the inner nebulosity. Several studies of this nebula 
have been published (e.g., Corradi et al. 1997), and it was
found that the bullets have low velocities -- but they may be moving nearly in
the plane of the sky. Our images show: (a) that there are distinct bow-shocks
around the bullets, indicative of outward motion (see Figure 2), and (b) that
one of the trails/jets leads back to a conical structure in the inner nebulosity.
What is curious is that, although the bullets that terminate the jets are aligned
with the star, the jet leading to the conical structure shows a pronounced lack
of alignment with the star. One possible explanation is that the jet which 
produced the trail out to the bullet has turned off, and subsequent motions of
the inner nebulosity have shifted the inner part of the trail.

The type of conical structure noted in the inner part of IC~4593, which is widest
near the star and narrows to an apex as we move away from the star, is also seen 
in other objects: M1-66, M1-92, and He~3-1475. The apex may be followed by an opening 
cone or outward spray. Such structures suggest flows that are being focused on a large
scale -- perhaps by oblique cooling shocks -- rather than jets emerging from stellar
accretion disks. The most spectacular object is the proto-PN He~3-1475 (Borkowski
et al. 1997). It was the first object observed in our program, and we have since
followed up with spectroscopic, infrared and polarization observations with Hubble's
STIS, NICMOS, and FOC instruments.

\begin{figure}  
\plotfiddle{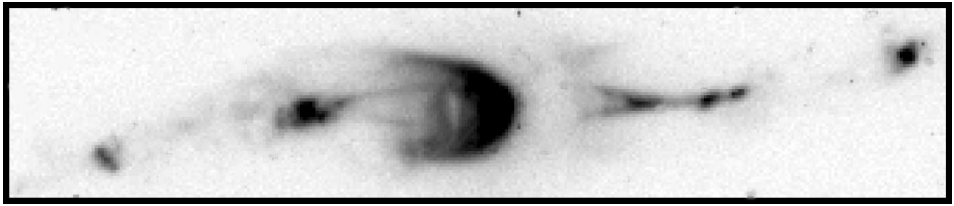}{2.8cm}{0.0}{70}{70}{-180}{-5}    
\caption{He 3-1475 through the HST [N II] filter.}
\end{figure}

The STIS results confirm the high velocities found by Riera et al. (1995). The
velocity increases down the axis of the approaching jet, reaches  
a maximum of $- 970$ km/s a bit before the cone apex, then declines
through the apex, followed by an abrupt deceleration of over 500 km/s when 
the flow hits the first knot. The flow in the receding jet is similar, with
a maximum velocity of $+ 895$ km/s. Though one would expect shocks at the knots
to produce very high temperatures, NICMOS images show H$_2$ emission from the 
knots. 

This work was supported by NASA through grant GO-06347.01-95A from STScI,
which is operated by AURA under NASA contract NAS5-26555.


\begin{references}

Borkowski, K.J., Blondin, J.M. \& Harrington, J.P. 1997, \apj, 482, L97.

Borkowski, K.J., Harrington, J.P. \& Tsvetanov, Z.I. 1995, \apj, 449, L143.

Corradi, R.L.M., Guerrero, M., Manchado, A. \& Mampaso, A. 1997, New Astronomy, 2, 461.

L\'{o}pez, J.A., Steffen, W., \& Meaburn, J. 1997. \apj, 485, 697.

Miranda, L.F. \& Solf, J. 1992, \aap, 260, 397.

Riera, A.A., Garcia-Lario, P., Manchado, A., Pottasch, S.R. \&
Raga, A.C. 1995, \aap, 302, 137.

\end{references}
\end{document}